\documentclass{IEEEtran}
\usepackage[T1]{fontenc}
\usepackage[utf8]{inputenc}
\usepackage{pgfplots}
\pgfplotsset{compat=newest}
\usepgfplotslibrary{patchplots}
\usepackage{dblfloatfix}
\usepackage{grffile}
\usepackage[cmex10]{amsmath}
\usepackage{mathtools}
\usepackage[standard]{ntheorem}
\usepackage{circuitikz}
\usepackage{tikz}
\usetikzlibrary{plotmarks}
\usetikzlibrary{arrows.meta}
\usetikzlibrary{decorations.pathmorphing}
\usetikzlibrary{shapes,arrows,positioning,calc}
\usetikzlibrary{circuits.ee.IEC}
\usetikzlibrary{arrows.meta, positioning}
\tikzset{every picture/.append style={font=\normalsize}}

\newcommand{\mc}{\mathcal}
\DeclareMathOperator{\diag}{diag}
\usepackage{ntheorem}
\DeclareMathOperator{\diff}{d}
\newcommand{\ddt}{\tfrac{\diff}{\diff \!t}}

\newtheorem{assumption}{Assumption}
\pgfplotsset{compat=newest}

\begin{document}
\title{Input-Output Specifications of Grid-Forming Functions and Data-Driven Verification Methods}

\author{Jennifer T. Bui and Dominic Gro\ss{} \thanks{This work was supported by the U.S. Department of Energy's Office of Energy Efficiency and Renewable Energy (EERE) under the Solar Energy Technologies Office Award Number 38637. The views expressed herein do not necessarily represent the views of the U.S. Department of Energy or the United States Government. J. T. Bui and D. Gro\ss{} is with the Department of Electrical and Computer Engineering at the University of Wisconsin-Madison, USA; e-mail:\{jtbui, dominic.gross\}@wisc.edu}}

\maketitle
\begin{abstract}
  This work investigates interoperability and performance specifications for converter interfaced generation (CIG) that can be verified using only input-output data. First, we develop decentralized conditions on frequency stability that account for network circuit dynamics and can be verified using CIG terminal dynamics and a few key network parameters. Next, we formalize performance specifications that impose requirements on the CIG disturbance response. A simple data-driven validation method is presented that enables verification of the interoperability and performance specifications for CIG using input-output data from a two-node system. Data obtained from electromagnetic transient (EMT) simulations are used to illustrate the proposed approach and the impact of key parameters such as inner control loop gains, network coupling strength, and controller bandwidth limitations.
\end{abstract}


%
\section{Introduction}
Electric power systems are undergoing an unprecedented transition towards large-scale deployment of renewable generation and energy storage interfaced by power electronics. Replacing conventional synchronous generators (SGs) with converter-interfaced generation (CIG) results in significantly different power system dynamics and challenges standard operating paradigms and controls on timescales from seasons to milliseconds. In the context of power system dynamics, the heterogeneous dynamics of renewables and CIG represent a significant barrier to their large-scale deployment. Specifically, scalable and reliable operation of today's systems crucially hinges on the largely homogeneous physics and controls of SGs \cite{PM2020}. In contrast, the dynamics of CIG vastly differ across technologies and controls resulting in interoperability concerns that are a major barrier to replacing centralized bulk power generation with a vast number of smaller distributed and heterogeneous resources.

Typically, control strategies for CIG are broadly categorized into (i) grid-following (GFL) controls that require a stable ac voltage at their point of interconnection (i.e., ensured by SGs), and (ii) grid-forming (GFM) controls that impose a stable ac voltage at their terminal and that self-synchronize are envisioned to be the cornerstone of future power systems \cite{gfmkeyetal}. Prevalent GFM controls such as droop control \cite{CDA1993}, virtual synchronous machine control \cite{DSF2015}, and dispatchable virtual oscillator control \cite{GCB+2019}, largely exhibit a similar response under realistic tuning \cite{gfmfreq} and can be tuned to coincide under simplifying assumptions \cite[Fig.~7]{DG2023}. While these results are promising, they do not establish formal interoperability guarantees of GFM controls and do not address concerns around interoperability with legacy generation and GFL controls. The operation of emerging power systems requires both GFM controls to replace functions of SGs and GFL controls to, e.g., enable maximum power point tracking (MPPT) for renewables and reliably operate HVDC transmission \cite{GSA+2021}. At the same time, a wide range of adverse interactions between SG controls and GFM controls \cite{CTG+2020}, SGs and GFM and GFL converters \cite{MSA+2021}, and GFM and GFL control \cite{DWT+2023} have been reported. 

A closely related problem is that, as of now, there is no clear or universally accepted definition of GFM control, and its distinction from GFL control continues to be a topic of discussion. The most prevalent method for broadly categorizing GFM and GFL control depends on whether the inverter is controlled as a voltage or current source \cite{acpower} and requires detailed models of the internal hardware and control structure. However, detailed models are often inaccessible due to intellectual property protections. In contrast, \cite{powertech} aims to characterize GFM controls through their ability to locally suppress frequency fluctuations. While this approach directly allows for experimental validation, it requires internal control signals, does not fully characterize standard CIG functions (e.g., $P-f$ droop), and does not address interoperability. Instead, from a system theoretic viewpoint, interoperability can be tackled through decentralized conditions that ensure small-signal frequency stability \cite{stability}. However, this method only considers a quasi-steady-state network model that does not adequately model adverse interactions between GFM converters and network circuit dynamics \cite{VHP+2017,GCB+2019,compensate}.

To address these challenges, this work combines decentralized interoperability and performance requirements with a data-driven verification method that only requires input-output data. We first extend the analytical small-signal stability conditions from \cite{stability} to account for network circuit dynamics. This results in conditions on the local CIG dynamics that can be graphically verified using a few key system parameters (i.e., lowest line inductance, R/X ratio) and transfer functions that relate the signals at the CIG terminal. Next, we illustrate that performance specifications discussed in the literature (see, e.g., \cite{UNIFIGFM23etal}) can be formalized as restrictions on the CIG terminal dynamics. Crucially, we do not aim to distinguish between GFM and GFL control but propose to formulate technology-agnostic performance requirements on the CIG disturbance rejection. This directly recovers the "frequency smoothing" capability used in \cite{powertech} to define GFM control. However, we also show that this ability to locally suppress frequency fluctuations does not sufficiently characterize common GFM functions such as $P-f$ droop. Finally, we develop a simple data-driven method that allows for experimental validation of the proposed interoperability and performance specifications. Specifically, the proposed approach leverages a two-bus system to obtain a frequency response model of a device under test (e.g., CIG) that can be used to verify our interoperability and performance specifications using a frequency gridding approach. The interoperability and performance specifications as well as the data-driven verification method are illustrated using data for standard GFM and GFL control obtained from detailed EMT simulations.

\section{Motivation and Problem Setup}
In the context of power systems, interoperability and the aforementioned information exchange has to be understood in the context of various time-scales. While control that leverages communication networks is commonly used on time-scales beyond seconds (e.g., secondary control and tertiary control) the information exchange on the real-time control layer (e.g., primary frequency control) happens through the power network. For example, synchronous generators self-synchronize their frequency through the power flows in the network (information exchange) and local controls use this fact by respond to frequency deviations (e.g., turbine governor system). Thus, synchronous generator dynamics and their controls are generally interoperable. However, for IBRs arbitrary dynamics can be imposed by their controls that do not necessarily synchronize through the grid. 

Thus, to ensure interoperability of IBRs, one needs to ensure that the local dynamics adhere to common specifications that enable (i) synchronization through the physics, and (ii) a coherent response that uses the information exchanged through the power network. In this sense, interoperability imposes restrictions on the dynamic interactions of IBRs through the power system in order to maintain the stability of the entire system as well as prevent destabilizing events from propagating. In this work, we specifically focus on the interoperability of frequency dynamics, that is, the stability of the relationship between ac voltage bus frequencies and active power flow.

\begin{figure}[h!]
  \centering
  \includegraphics[width = 0.9\columnwidth]{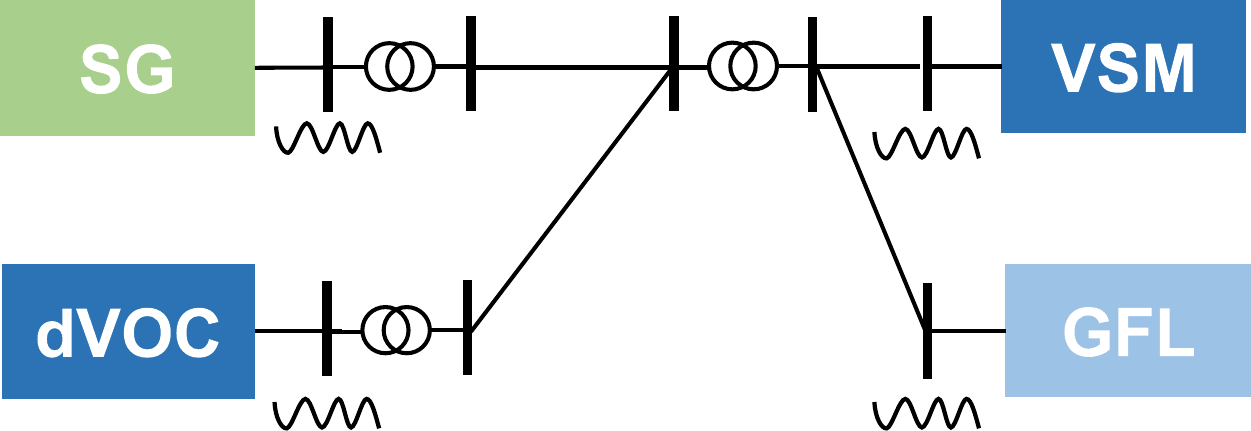}
  \vspace*{3mm}
  \caption{Power system consisting of generation units with synchronous generators and converters with different controls.}
  \label{fig:original-network}
\end{figure}

In the context of a power system with a mix of converter-interfaced and conventional thermal generation, such as the one depicted in Figure~\ref{fig:original-network}, it is important to consider the compatibility of the IBR different control schemes and legacy generators under interconnection. To develop specifications on the local IBR bus dynamics, we separate the problem into two steps. To this end, we require a network model that abstractly captures how the network circuit itself impacts the exchange of signals that are then received as inputs by the connected devices, i.e., to capture the impact of network topologies and line parameters through a few key parameters. In a second step, we will analyze how the local device outputs (i.e., frequency and voltage magnitude) respond to its inputs (i.e., network power injections) and determine suitable bounds on these dynamics to ensure interoperability for a given set of (abstract) network parameters.

As we will show in the remainder of this manuscript, the system properties that are relevant to the interoperability and stability of the system-wide frequency dynamics include the $X/R$ ratio, line and transformer inductance, and topology. These properties help determine connectivity and coupling strength, attributes that impact power flow across the system. However, for various reasons such as scalability and robustness, it is not desirable to certify stability only for a specific network topology and inductances. Therefore, key salient features of the network that influence stability have to be identified that allow to certify stability independently of the network topology and precise line parameters. As our analysis in the subsequent section shows, these key salient parameters of the network are the $X/R$ ratio as well as strongest and weakest coupling in the Kron-reduced network. Broadly speaking, the strongest and weakest coupling can be understood as the largest and smallest short-circuit ratio or the smallest and largest inductance between devices when eliminating network buses without generators.

\begin{figure}[h!!!]
  \centering
  \includegraphics[width = 0.9\columnwidth]{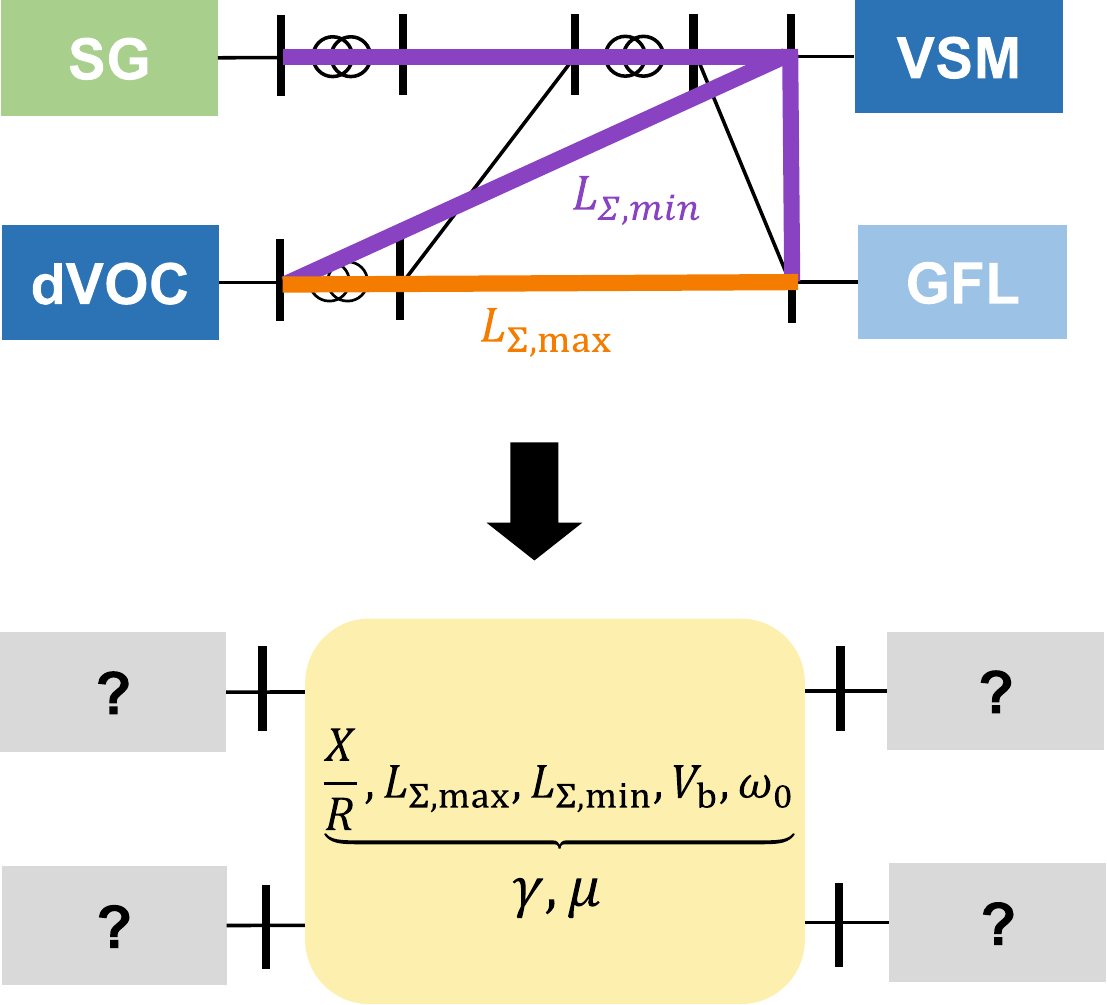}
  \vspace*{3mm}
  \caption{Abstracting the original network (top) through an analytical model (bottom) parameterized in key network parameters.  \label{fig:abstracted-network}}
\end{figure}

With this information, for the purpose of developing analytical stability certificates, it is possible to abstractly model a network such as Figure~\ref{fig:original-network} through the abstract parameters illustrated in Figure~\ref{fig:abstracted-network} where connectivity and coupling strength are encoded through the scalar parameter $\gamma$ and $\mu$ using $L_{\Sigma, max}$ and $L_{\Sigma, max}$ (the highest and lowest outgoing impedance of interconnection buses), $X/R$ ratio, nominal bus voltage magnitude $V_b$, and the nominal frequency $\omega_0$. Please see the following section for further details.

Moreover, in this work, the dynamic response of devices in the network is not modeled using white box models of their control or hardware implementation since this information is typically inaccessible. Instead, we leverage data-driven input-output models to model the dynamic behavior of generating units (i.e., IBRs, SGs, etc.). Combining the models of the local device dynamics (e.g., at the bus of interconnection) with that of the abstracted network, it is possible to evaluate interoperability, stability, and performance.  By imposing a decentralized stability condition, the local device dynamics are restricted in such a way that maintains stability across the network. Given a few key network parameters this approach allows to test each device in isolation to (i) verify if it meets interoperability and performance requirements for the abstract network parameters, and (ii) understand the impact of the abstract network parameters. Stability and interoperability are then guaranteed for arbitrary topologies and line parameters as long as the $X/R$ ratio, the strongest coupling, and the weakest coupling do not change.

\begin{figure}[h!!]
  \centering
  \includegraphics[width = 0.9\columnwidth]{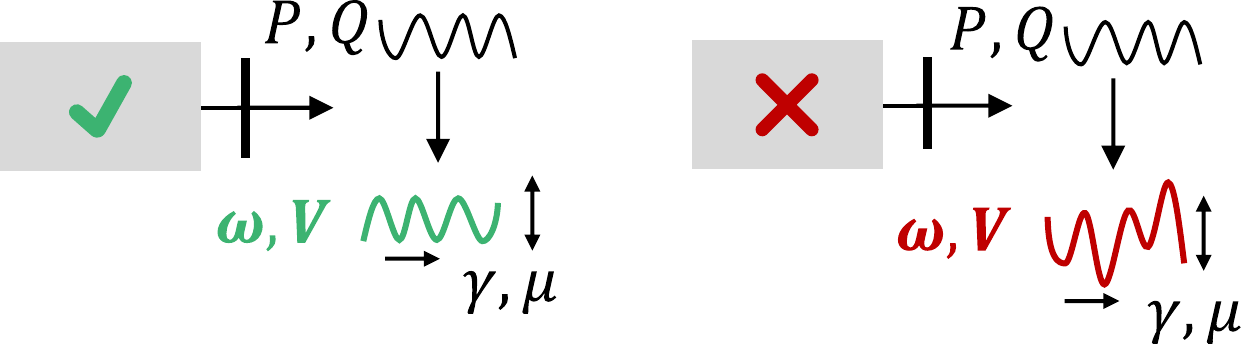}
  \vspace*{3mm}
  \caption{Verification of a decentralized stability condition for each individual device.}
  \label{fig:decentralized-condition}
\end{figure}

Figure~\ref{fig:decentralized-condition} illustrates this approach in which the dynamic response of bus voltages (i.e., frequency and magnitude) to network power injections (e.g., active and reactive power) is analyzed and compared to specifications parameterized in the abstract network parameters.  If the local dynamic response is within prescribed bounds, interoperability can be guaranteed. If the local dynamic response is outside the prescribed bounds, then interoperability cannot be guaranteed through our decentralized stability condition. In other word, the system may still be stable even though it does not pass the decentralized stability conditions. However, the lack of a-priori guarantees that are largely independent of the precise network parameters results in scalability concerns in this case. Therefore, using decentralized stability conditions we can certify small-signal stability using only a few key network parameters and the individual local dynamics. In contrast, when this approach fails the system may be unstable or certifying stability may require exact network and device models and arbitrary many simulations to certify stability. In other words, the second approach is not scalable or amenable to developing standards or grid codes. In contrast, the decentralized stability certificates can directly inform standards and grid codes to ensure and verify interoperability for a wide range of system configurations with a single test of each device.

\vspace{2mm}

There are two key advantages to this approach. The first advantage is its independence from detailed models of both the network and connected devices.  Instead, the proposed methods only utilize commonly known system properties and measured data from a simple experimental test bed. The second benefit of this framework is that it allows for the mapping of system-level needs to device-level specifications. Manufacturers are encouraged to implement the proposed methods themselves when they test their devices. They can adjust their designs accordingly to meet requirements outlined by grid operators, improving the effectiveness and efficiency of deployment. 

Interoperability does not guarantee performance (e.g., specifically desired functionality/capability). Thus, this work also investigates the development of performance specifications that can be verified using input-output data. First, we obtained reduced-order models of generic converter controls. Then, we derived several transfer functions that correspond to accessible parameters (i.e., power injection, bus voltages). We analyzed the transfer functions, including graphically, to identify salient traits that can differentiate device functions. Finally, we conducted a numerical study in which we used data collected from detailed EMT simulations to recover the transfer functions previously derived and compared the analytical and numerical results.

\section{Power system model}

\subsection{Network topology}
We model the power system as an undirected graph $\mc G=(\mc N,\mc E)$ with nodes $\mc N$ corresponding to $|\mc N|$ buses and edges $\mc E \subseteq \mc N \times \mc N$ corresponding to $|\mc E|$ transmission lines. Here, $|\mc X|$ denotes the cardinality of a set $\mc X \subset \mathbb{N}$. To every bus $n \in \mc N$, we associate a voltage phase angle $\theta_n$, frequency $\omega_n=\ddt \theta_n$, voltage magnitude $V_n$, and active and reactive power injections $p_n$ and $q_n$. Moreover, to every line $m \in \{1,\ldots,|\mc E|\}$ connecting buses $n_m$ and $k_m$ we associate an active and reactive power flow $p_{n_m,k_m}$ and $q_{n_m,k_m}$. Thus, the bus power injections are given by $p_n= \sum_{(n,k) \in \mathcal{E}}  p_{n,k}$ and $q_n= \sum_{(n,k) \in \mathcal{E}} q_{n,k}$. In the remainder, we encode the network topology using the oriented incidence matrix $B \in \mathbb{R}^{|\mc N| \times |\mc E|}$ (see \cite{compensate}). 

\subsection{Device model}
\begin{figure}[b!]
  \centering
  \includegraphics[width=0.6\columnwidth]{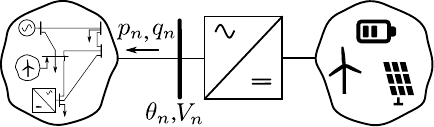}
  \caption{Converter-interfaced generation connected to a grid. \label{fig:CIG}}
\end{figure}

To obtain a technology- and control-agnostic model, we model the dynamic response of generation devices at every bus $n \in \mc N$ (see Fig.~\ref{fig:CIG}) through the transfer function model
\begin{align}\label{eq:busdyn}
  \begin{bmatrix} \omega_n (s) \\ V_n(s) \end{bmatrix} =
  \begin{bmatrix}
    g_{\omega_n,p_n}(s) & g_{\omega_n,q_n}(s) \\ g_{V_n,p_n}(s) & g_{V_n,q_n}(s)
  \end{bmatrix}
  \begin{bmatrix} p_n(s) \\ q_n(s) \end{bmatrix}.
\end{align}

We emphasize that the choice of inputs and outputs in \eqref{eq:busdyn} is somewhat arbitrary. For example, GFL control using a synchronous reference frame phase-locked loop (SRF-PLL) controls the power injection $p_n$ and $q_n$ as a function of $\omega_n$ and $V_n$, while a GFM converter or SG controls $\omega_n$ and $V_n$ as a function of $p_n$ and $q_n$. Nonetheless, both cases can be represented through the model \eqref{eq:busdyn} that simply encodes the dynamic relationship between the signals $\omega_n$, $V_n$ and $p_n$, $q_n$ (see Sec.~\ref{ex:gfmdroop} and Sec.~\ref{ex:gfl}).

\subsection{Transmission line model}
It remains to model the network. Analogously to the bus model \eqref{eq:busdyn}, we model the transmission line $m \in \{1,\ldots,|\mc E|\}$ connecting buses $(n_m,k_m) \in \mc E$ using the model
\begin{align*}
  \begin{bmatrix} p_{n_m,k_m} \\ q_{n_m,k_m} \end{bmatrix} =
  \begin{bmatrix}
    g_{p_m,\theta_m}(s) & g_{q_m,\theta_m}(s) \\ g_{p_m,V_m}(s) & g_{q_m,V_m}(s)
  \end{bmatrix}
  \begin{bmatrix} \theta_{n_m}(s)-\theta_{k_m}(s) \\ V_{n_m}(s)-V_{k_m} \end{bmatrix}.
\end{align*}
%
%
For brevity of the presentation, we will focus on frequency dynamics and model the dynamics of the $|\mc E|$ transmission lines using the second-order transfer function 
\begin{align}
  g_{p_m,\theta_m}(s) \coloneqq \underbrace{\frac{\omega_0 V^\star_{n_m} V^\star_{k_m}}{\ell_m}}_{\eqqcolon \kappa_m} \underbrace{\frac{1}{s^2+2 \rho_m s+\omega_0^2+\rho^2_m}}_{\eqqcolon \mu_m(s)},
\end{align}
with nominal bus voltages $V^\star_{n}\in\mathbb{R}_{>0}$,resistance-inductance ratio $\rho_m=\frac{r_m}{l_m} \in \mathbb{R}_{>0}$. We emphasize that this model generalizes the commonly used quasi-steady-state network model $g_{p_m,\theta_m}(0)$ to capture network circuit dynamics, up to approximately line frequency, that play a crucial role in small-signal frequency stability analysis of converter-dominated power systems \cite{GCB+2019,compensate}. 

\subsection{Multi-converter / multi-machine network model}
To simplify the overall network model, we require the following assumption.

\begin{figure}[b!!]
  \centering
  \includegraphics[width=0.7\columnwidth]{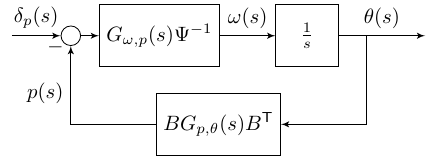}
  \caption{Small-signal frequency dynamics.\label{fig:smallsignal}}
\end{figure}

\begin{assumption}{\bf(Decoupled active and reactive power)}\label{ass.coupling}
  We assume that active and reactive power are decoupled, i.e., $g_{q_m,\theta_m}(s)=g_{p_m,V_m}(s)=0$ for all $m \in \{1,\ldots,|\mc E|\}$.
\end{assumption}
The vector of bus power injections $p \in \mathbb{R}^{|\mc N|}$ is given by $p = B G_{\omega,p}(s) B^\mathsf{T} \theta$. To analyze the impact of device ratings, we use $\psi_n \in \mathbb{R}_{>0}$ to denote the device rating at every bus $n \in \mc N$. Moreover, we define the matrix  $G_{\omega,p}(s) \coloneqq\diag\{ g_{\omega_n,p_n}(s)\}_{n=1}^{|\mc N|}$ of bus transfer functions $g_{\omega_n,p_n}(s)$ normalized by the device rating $\psi_n$ and the matrix $\Psi \coloneqq \diag\{\psi\}_{n=1}^{|\mc N|}$ collecting the device ratings. Finally, to compactly represent our model, we define the matrix of transmission line transfer functions $G_{p,\theta}(s) \coloneqq \diag\{g_{p_m,\theta_m}(s)\}_{m=1}^{|\mc E|}$.
The small-signal frequency dynamics are shown in Fig.~\ref{fig:smallsignal}. We emphasize that this model can capture the frequency dynamics of a wide range of devices from GFM converters and GFL converters to SGs. 

To simplify the analysis, we make the following assumption that is typically justified for lines at the same voltage level.
\begin{assumption}{\bf(Uniform resistance-inductance ratio)}\label{ass.constant.ratio}
The resistance-inductance ratio is identical for all lines, i.e., $\rho_m=\rho$ for all $m \in \{1,\ldots,|\mc E|\}$.
\end{assumption}
Under Assumption~\ref{ass.constant.ratio}, $\mu(s)=\mu_m(s)$ for all $m \in \{1,\ldots,|\mc E|\}$. Letting $K \coloneqq \diag\{\kappa_m\}_{m=1}^{|\mc E|}$, the network circuit dynamics simplify to $B G_{p,\theta}(s) B^\mathsf{T}=\mu(s) B K B^\mathsf{T}$, where $L \coloneqq B K B^\mathsf{T}$ is the graph Laplacian of the network.

In the remainder, we aim to develop decentralized conditions for stability under interconnection and local performance specifications on the transfer functions in \eqref{eq:busdyn}. Subsequently, we show how these specifications can be verified using only input-output data and simple graphical methods.

\section{Interoperability \& Performance Specifications}
%
%
\subsection{Interoperability of Frequency Dynamics}
We first focus on interoperability of bus frequency dynamics. Specifically, we define bus frequency dynamics as interoperable if they meet decentralized stability conditions that ensure stability under interconnection and extend a decentralized stability criterion from \cite{stability} to account for network circuit dynamics. 

\begin{figure}[b!!] 
  \centering
  \includegraphics[width=0.7\columnwidth]{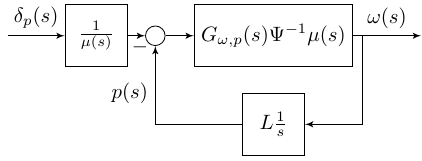}
  \caption{Rearranged small-signal frequency dynamics with constant resistance-inductance ratio.\label{fig:smallsignalre}}
\end{figure}

To state our decentralized stability condition, we define the constant $\gamma_n \coloneqq 2 \sum_{(n,k) \in \mc E} \frac{\omega_0}{\ell_{n,k}} V_n V_k$ where, with a slight abuse of notation, $\ell_{n,k}\in\mathbb{R}_{>0}$ denotes the inductance of the line connecting bus $n$ and $k$. This constant models the key parameters of the system that influence frequency stability. In particular, $\gamma_n$ can be bounded by $\gamma \coloneqq 2  \frac{e_{\max}}{\ell_{\min}} \omega_0 V^2_{\max}$, where $e_{\max} \in \mathbb{N}$ is an upper bound on the number of outgoing edges of all nodes, $\ell_{\min} \in \mathbb{R}_{>0}$ is a lower bound on the line impedances, and maximum voltage magnitude $V_{\max}$.
\begin{theorem}{\bf(Decentralized stability condition)}\label{th.stability}
  Assume that all poles of $g_{\omega_n,p_n}(s)$ are in the open left half-plane for all $n \in \mc N$. The overall frequency dynamics shown in Fig.~\ref{fig:smallsignalre} are asymptotically stable if there exists $\alpha \in [0,\pi/2)$ such that 
  \[ \operatorname{Re}\left\{e^{j\alpha}(1+\tfrac{\gamma}{\psi_n} \tfrac{\mu(j\omega_p)}{j \omega_p} g_{\omega_n,p_n}(j\omega_p))\right\}>0\]
  holds for all $n \in \mc N$ and $\omega_p \in \mathbb{R} \cup \{\infty\}$.
\end{theorem}
The proof of the Theorem follows by applying elementary block operations to obtain the block diagram shown in Fig.~\ref{fig:smallsignalre}. Because $\mu(s)^{-1}$ is asymptotically stable, it remains to show that the closed loop with $\delta_p(s)=0$ is stable. This directly follows from \cite[Lem.~5, Th.~1]{stability}, scaling $L$ as in \cite[Fig.~4]{stability}, and noting that $\gamma$ is an upper bound on the largest eigenvalue of the graph Laplacian $L$, and applying the frequency response bounds in  \cite[Sec.~III-C]{stability}. 

Notably, the stability condition of Theorem~\ref{th.stability} account for key system parameters such as the R/X ratio and coupling strength (i.e., lowest line inductance), but does not require knowledge of the exact line parameters or network topology. In particular, conditions can be applied for each bus without assembling the overall network model. Moreover, the conditions of Theorem~\ref{th.stability} can be checked graphically through Nyquist plots as discussed in Sec.~\ref{ex:interoplinedyn} and Sec.~\ref{sec:numintop}.

While Theorem~\ref{th.stability} allows certification of small-signal frequency stability under interconnection, it does not provide any performance guarantees (e.g., frequency damping). Instead, the next section focuses on local performance specifications.

%
\subsection{Performance Specifications}\label{sec:performance}
\begin{figure}[b]
  \centering
  \includegraphics[width=0.8\columnwidth]{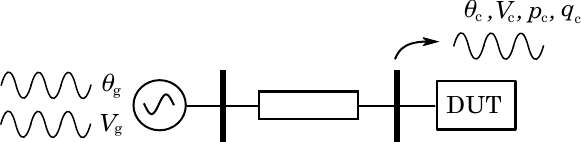}
  \caption{Two-bus system used to define and assess performance specifications. 
  \label{fig:testbench}}
\end{figure}

Local performance specifications are developed that restrict the response of CIG to disturbances. We will first mathematically define requirements and, subsequently, provide illustrative examples. To this end, we use the two-bus system shown in Fig.~\ref{fig:testbench}, consider $\omega_g$ and $V_g$ as disturbance inputs, and formulate performance specifications on the ability of the device under test (DUT) to reject grid disturbances at its local bus.
We first consider the quasi-steady-state network model to characterize the desired response and then verify the specifications for typical controls in Sec.~\ref{sec:bodevalid} using EMT simulations. The response of the CIG bus frequency to a grid frequency perturbation $\omega_g$ is given by 
\begin{align}
  g_{\omega,\omega_g}(s) = \frac{g_{p,\theta}(0)g_{\omega,p}(s)}{s + g_{p,\theta}(0) g_{\omega,p}(s)}.
\end{align}
Next, let  $\omega_p$ denote the frequency of the perturbation applied to the grid frequency $\omega_g$. Then, 
  \begin{align}\label{eq:upperbound}
    |g_{\omega,\omega_g}(j\omega_p)| \le \left|\frac{1}{\tau_fj\omega_p+1}\right|
  \end{align}
  ensures a well-damped local CIG bus frequency at high frequencies and rules out amplification of low-frequency oscillations. This condition combines the "frequency smoothing" requirement proposed in \cite{powertech} with a passivity requirement at low frequencies. In other words, the Bode magnitude plot is restricted to a maximum of unity gain at low frequencies,  indicating passive behavior that prevents resonance under all frequencies. Moreover, the unit dc gain implies synchronization with the infinite bus.  A high-frequency roll-off follows afterward, which begins at a corner frequency determined by $\tau_f$. Thus, condition \eqref{eq:upperbound} imposes conditions on the transient stability of a CIG, requiring it to both prevent amplification of and attenuate high-frequency disturbances.
With transfer function $g_{p,\omega_g}(s)$, we can characterize droop functions for active power and frequency using only input-output data, i.e., without requiring a specific control implementation.  Given an active power droop gain $m_p \in \mathbb{R}_{>0}$, a tolerance $\epsilon \in \mathbb{R}_{>0}$, and frequency range $\omega_p \in [0,\bar{\omega}]$ on which the CIG should provide frequency droop, results in the specification
\begin{align}\label{eq:droopbound}
  m^{-1}_p - \epsilon \leq |g_{p,\omega_g}(j\omega_p)| \le m^{-1}_p + \epsilon, \quad \forall \omega_p\leq \bar{\omega}.
\end{align}
We emphasize that \eqref{eq:upperbound} and \eqref{eq:droopbound} are distinct specifications and controls may satisfy \eqref{eq:upperbound} without providing the frequency droop response specified by \eqref{eq:droopbound} (see Sec.~\ref{ex:gfmpi}).

\subsection{Example: GFM Droop control}\label{ex:gfmdroop}    The most prevalent form of GFM control is droop control, which imposes the linear relationship 
\begin{align}        \label{eqn:gfm_c}
   g_{\text{droop}}(s) = \frac{m_p}{\tau s + 1}
\end{align}
between active power injection as the input and frequency as the output, where $\tau \in \mathbb{R}_{>0}$ is the time constant for filtering active power measurements.  

    \begin{figure}[t]
      \newlength{\figW}
      \newlength{\figH}
      \setlength\figH{0.4\linewidth}
      \setlength\figW{0.82\linewidth}
      \centering
      \input{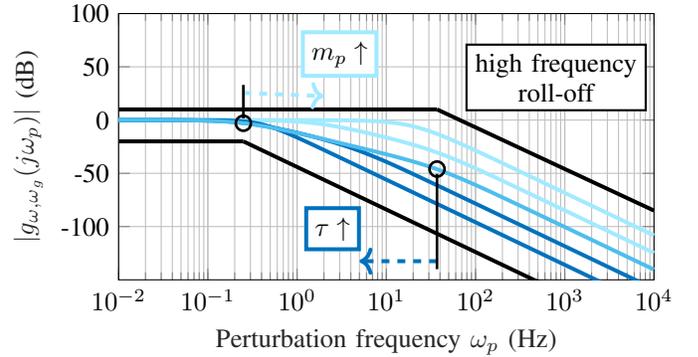}
      \caption{Bode magnitude plot $g_{\omega,\omega_g}(s)$ for GFM droop control. The droop constant $m_p$ shifts the first corner frequency. The time constant $\tau$ shifts the second pole and the start of the high-frequency roll-off.      \label{fig:gfm_specs}}
    \end{figure}

  The Bode magnitude plot of $g_{\omega,\omega_g}(s)$ with $g_{\omega,p}(s)=g_{\text{droop}}(s)$ is shown in Fig.~\ref{fig:gfm_specs}. For low frequency oscillations, the CIG does not damp the CIG bus frequency but synchronizes with the infinite bus.  Once oscillations surpass the cut-off frequency, the gain decreases, i.e., illustrating that droop control achieves high-frequency damping. The dependency of $g_{\omega,\omega_g}(s)$ on the control variables $m_p$ and $\tau$ is shown in Fig.~\ref{fig:gfm_specs}.

  \subsection{Example: SRF-PLL GFL control}\label{ex:gfl}
    Ubiquitous implementations of GFL control utilizes a SRF-PLL and frequency damping.  The active power injection of a GFL CIG as a function of frequency is given by
    \begin{subequations}\label{eqn:gfl_c}
      \begin{align} 
        g_{\text{gfl}}(s) & = g^{-1}_{PLL}(s)g^{-1}_D(s) \\
        & = \frac{s^2 + k_ps + k_i}{k_ps + k_i} \; \frac{\tau_ds + 1}{D}
      \end{align}
    \end{subequations}
      where $k_p \in \mathbb{R}_{>0}$ and $k_i \in \mathbb{R}_{>0}$ are the PLL gains and $\tau_d \in \mathbb{R}_{>0}$ is the time constant of the realizable differentiator for implementing frequency damping.

    \begin{figure}[b]
      \setlength\figH{0.35\linewidth}
      \setlength\figW{0.82\linewidth}
      \centering
      \input{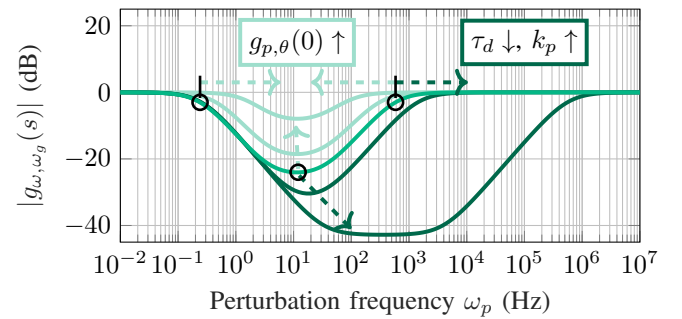}
      \caption{Bode magnitude plot of $g_{\omega,\omega_g}(s)$ for SRF-PLL GFL control. Notably, SRF-PLL GFL control does not provide damping for low or high frequencies.
      \label{fig:gfl_specs}}
    \end{figure}

Fig.~\ref{fig:gfl_specs} shows the Bode magnitude plot of $g_{\omega,\omega_g}(s)$ for $g_{\omega,p}(s)=g_{\text{gfl}}(s)$.  It can be seen that SRF-PLL GFL control cannot achieve a high-frequency roll-off. 

\begin{proposition}{\bf(SRF-PLL damping)}\label{prop.SRFPLL}
There exist no coefficients $k_p \in \mathbb{R}_{>0}$, $k_i \in \mathbb{R}_{>0}$, $\tau_d \in \mathbb{R}_{>0}$, such that the SRF-PLL control \eqref{eqn:gfl_c} satisfies the performance specification \eqref{eq:upperbound}.
\end{proposition}
\begin{proof}
  Let $g_{\omega,p}(s)=g_{\text{gfl}}(s)$. Then, the relative degree of $g_{\omega,\omega_g}(s)$ is zero and it follows that $g_{\omega,\omega_g}(j \omega_p)$ converges to a constant as $\omega_p \to \infty$. This immediately contradicts \eqref{eq:upperbound}. \hfill $\blacksquare$
\end{proof}

Figure~\ref{fig:gfl_specs} shows two cut-off frequencies that define the range of damping. Because the GFL control utilizes a PLL, the stiffness of the network connection plays an important role in the converter's dynamic behavior. As the strength of the network connection increases (i.e., $g_{p,\theta}(0)\to \infty$), the nadir increases and the cut-off frequencies shift toward each other. In turn, the range and impact of the converter's oscillation damping decreases. More aggressive tuning (i.e., $\tau_d \to 0$ and $k_p \to \infty$) increases the second cut-off frequency to extend the range of damping and decrease the nadir, which improves performance. However, a stiffer network, i.e., a high short circuit ratio (SCR), will negate these effects and narrow the range of damping. Notably, practical bandwidth limits preclude highly aggressive tuning, i.e., imposing a lower bound on $\tau_d$ and upper bound on $k_p$, which restricts the maximum value of the second cut-off frequency.

 \subsection{Example: GFM PI control}\label{ex:gfmpi}
\begin{figure}[b]
    \setlength\figH{0.7\linewidth}
    \setlength\figW{0.82\linewidth}
    \centering
    \input{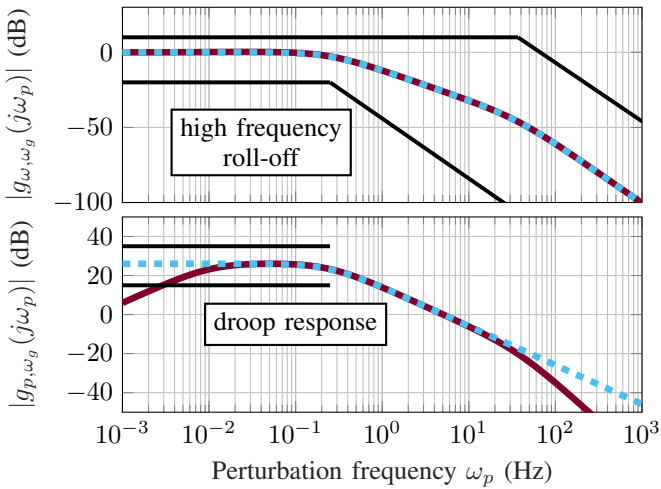}
    \caption{The magnitude Bode plots of $g_{\omega,\omega_g}(s)$ (top) for GFM droop \eqref{gfm_tf} and GFMPI \eqref{gfmpi_tf} controls look identical. However, the two controls exhibit different responses for $g_{p,\omega_g}(s)$ (bottom). GFM PI control does not have a steady-state droop response but displays high-frequency roll-off.}
    \label{fig:gfmpi_specs}
  \end{figure}

GFM control without active power droop can be implemented via the proportional-integral (PI) control 
\begin{align}\label{eq:gfmpi}
  g_{\text{gfm pi}}(s) = \frac{\xi_i s + \xi_p}{\tau s^2 + s}
\end{align}
  where $\xi_p$ and $\xi_i$ are the PI control gains. Note that \eqref{eq:gfmpi} has a pole at zero and does not satisfy the requirements of Theorem~\ref{th.stability}. This control is included here to illustrate the need to investigate performance specifications beyond the transfer function $g_{\omega,\omega_g}(s)$ considered \cite{powertech}. The GFM PI control can be designed to achieve the same response for $|g_{\omega_g, \omega_c}(s)|$ as GFM droop control.  The first plot in Fig.~\ref{fig:gfmpi_specs} illustrates that both controls have identical Bode magnitude plots.  
  However, GFM PI control can be differentiated from the GFM droop control when examining power injection instead of converter bus frequency.  The second plot in Fig.~\ref{fig:gfmpi_specs} demonstrates that the two controls differ in the low and high-frequency ranges.  In the low-frequency range, the active power injection from GFM droop control is dictated by the droop coefficient.  This feature is not present in the GFM PI response. highlighting that the high-frequency roll-off \eqref{eq:upperbound} discussed in \cite{powertech} does not fully characterize the performance and functionalities of CIG.

\subsection{Example: Interoperability with steady-state  network model}\label{ex:nyquist_alg}
\begin{figure}[ht]
    \setlength\figH{0.3\columnwidth}
    \setlength\figW{0.82\columnwidth}
    \centering
    \input{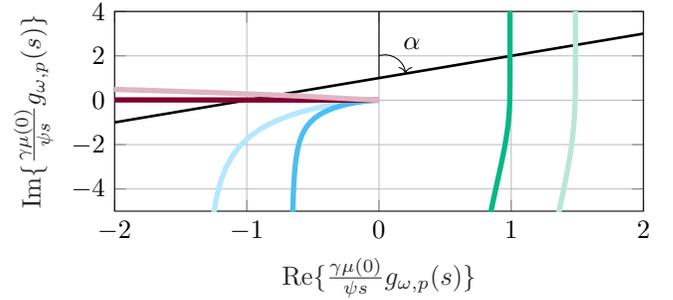}
    \caption{Nyquist plots of $\frac{\gamma \mu(0)}{\psi s} g_{p_c,\omega_c}(s)$ using an algebraic model of a static network for GFM droop (\ref{leg:gfm_ny}), GFM PI (\ref{leg:gfmpi_ny}), and SRF-PLL GFL control (\ref{leg:gfl_ny}).  The lighter curves show the Nyquist plots for increased $\frac{\gamma \mu(0)}{\psi}$, i.e., decreased minimum line inductance $\ell_{\min}$ or  rating $\psi$.
    \label{fig:nyquist_alg}}
  \end{figure}

We first illustrate the results of Theorem~\ref{th.stability} using the standard quasi-steady-state network model (i.e., $\mu(0)$) considered in \cite{stability}. The interoperability condition of Theorem~\ref{th.stability}, can be checked graphically by verifying that the Nyquist plot of $\tfrac{\gamma \mu(0)}{\psi s} g_{\omega,p}(s)$ is bounded by a half-plane crossing $(-1,0)$ with angle $\alpha$. The Nyquist plot of the controls considered in this section are shown in Fig.~\ref{fig:nyquist_alg}.
Since $g_{\text{droop}}(s)/s$ has only two poles and a maximum gain of unity, the Nyquist curve never crosses the imaginary or real axis while approaching the origin as $s \rightarrow \infty$. Therefore, we can always increase $\alpha$ to enclose the Nyquist curve of GFM droop control and guarantee stability and interoperability of systems that only contain GFM droop control for any network coupling strength or CIG rating. We must note that this result is misleading, as shown in \cite{compensate} and discussed in the next example. In particular, GFM control becomes unstable if the network coupling $\gamma$ is too strong or the device rating $\psi$ becomes too low.

On the other hand, Theorem~\ref{th.stability} cannot be used to certify stability of GFL and GFM for any network coupling strength $\gamma$. Specifically, as $\omega_p \rightarrow \infty$, the GFL Nyquist curve tends toward $\infty$ and cannot be bounded jointly with the GFM Nyquist curve as required by Theorem~\ref{th.stability}. We emphasize that the Nyquist plot of GFL CIG violates the stability bound for high frequencies at which the steady-state network model is not applicable.

GFM PI control has a pole at zero and, therefore, Theorem~\ref{th.stability} is not applicable. The Nyquist plot of GFM PI control is included for completeness. Note that it starts outside the stability bound and reaches within the boundary as $s \rightarrow \infty$. 

\subsection{Example: Interoperability with network circuit dynamics}\label{ex:interoplinedyn}
\begin{figure}[ht]
    \setlength\figH{0.3\columnwidth}
    \setlength\figW{0.78\columnwidth}
    \centering
    \input{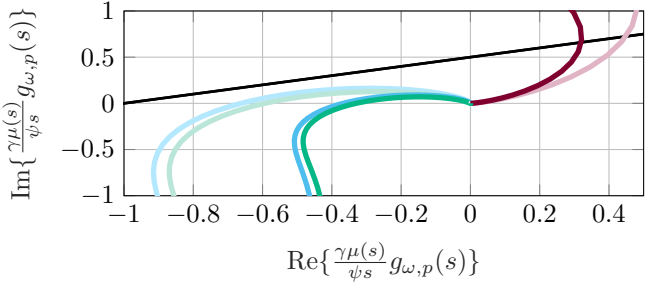}
    \caption{Nyquist diagram of $\tfrac{\gamma \mu(s)}{\psi s} g_{\omega,p}(s)$ for GFM droop (\ref{leg:gfm_ny}), GFM PI (\ref{leg:gfmpi_ny}), and SRF-PLL GFL (\ref{leg:gfl_ny}) control considering  network circuit dynamics.  The lighter curves show the Nyquist plots for increased $\frac{\gamma \mu(0)}{\psi}$, i.e., decreased minimum line inductance $\ell_{\min}$ or  rating $\psi$. \label{fig:nyquist_netd_tf}}
  \end{figure}

Figure~\ref{fig:nyquist_netd_tf} shows the Nyquist diagram of 
$\tfrac{\gamma \mu(s)}{\psi s} g_{\omega,p}(s)$ for various controls. 
A striking outcome is the similarity between the GFM droop and GFL curves. Loosely speaking, GFM droop and SRF-PLL GFL control exhibit a similar response in the low-frequency range and deviate for high frequencies. However, $\mu(s)$ is a low pass filter that attenuates the differences at high frequencies and hence the Nyquist plots of $\tfrac{\gamma \mu(s)}{\psi s} g_{\omega,p}(s)$ for both controls become similar.  

As the network connectivity increases, the margin between the Nyquist curves $\tfrac{\gamma \mu(s)}{\psi s} g_{\omega,p}(s)$ and the half-plane tightens for both GFM droop control and SRF-PLL GFL control.  Figure~\ref{fig:nyquist_netd_tf} reveals that stronger network coupling  (i.e., increasing $\gamma$) and reduced rating (i.e., decreasing $\psi$) eventually result in instability. This is in line with results for GFM in \cite{VHP+2017,GCB+2019,compensate} but, at first glance, may seem to contradict results that show that GFL control is vulnerable to instability under weak grid coupling \cite[Sec.~III-B]{LGG2022}. However, we emphasize that these results typically only apply to GFL CIG connected to an infinite bus, whereas in a multi-converter system, GFL instabilities under weak coupling may be related to voltage stability \cite[Sec.~VI-A]{LGG2022}. Overall, under our modeling assumptions, weak grid coupling affects GFL performance (see Sec.~\ref{ex:gfl}) but not necessarily frequency stability.

Finally, GFM PI control again violates the conditions of Theorem~\ref{th.stability}. Extending Theorem~\ref{th.stability} to account for GFM PI control is seen as interesting topic for future work.

\section{Data-Driven Validation}\label{sec:validation}
Finally, we develop a method for validating the proposed interoperability and performance specifications and apply it to input-output data obtained from detailed EMT simulations.

\subsection{Frequency response models from input-output data}\label{sec:sysid}
\begin{figure}[b!!]
  \centering
  \includegraphics[width=0.95\columnwidth]{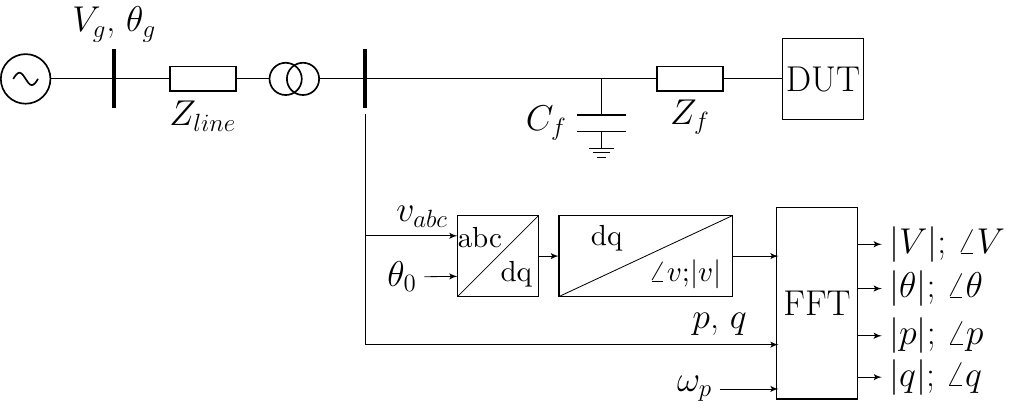}
  \caption{Two-bus system and signal processing for capturing frequency response models.}
  \label{fig:experiment}
\end{figure}

Conceptually, standard system identification methods could be used to obtain the bus transfer functions \eqref{eq:busdyn}. However, these typically fit a model with prescribed numbers of poles and zeros to the data. Crucially, the specifications developed in the previous sections do not require knowledge of the transfer functions in closed form but can be tested if the response at discrete frequencies is known. To this end, we use the two-bus system shown in Fig.~\ref{fig:experiment} that consists of a controlled ac voltage source and the DUT. 
  To recover a frequency response model, we follow a three-step procedure.
  \begin{figure*}[t]
    \setlength\figH{0.33\linewidth}
    \setlength\figW{0.9\linewidth}
    \centering  
    \input{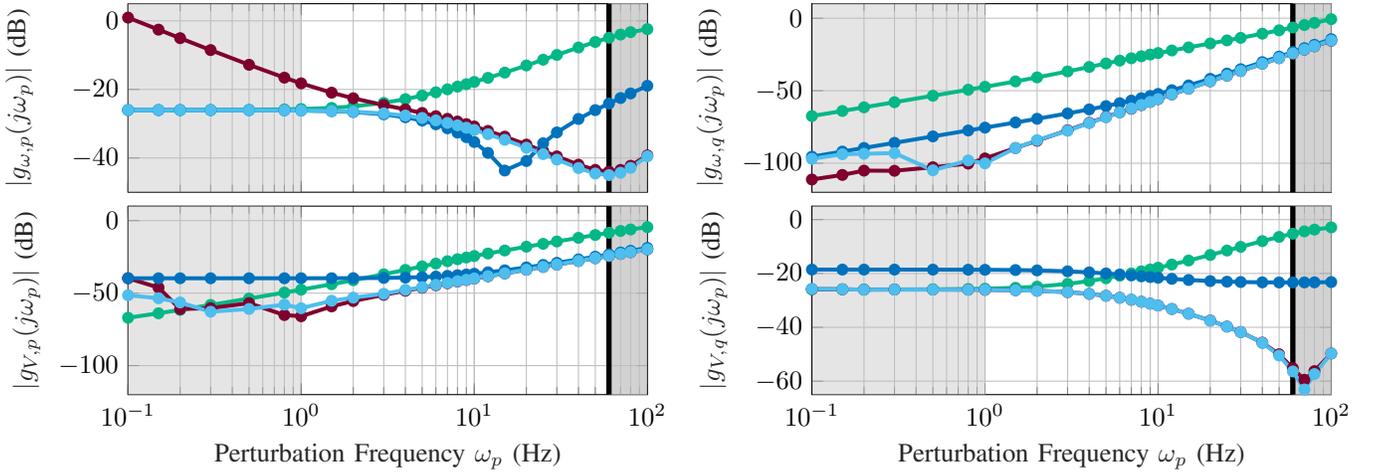}
    \caption{Bode magnitude plots that illustrate the local CIG dynamics for GFM single-loop droop (\ref{gfm_single_loop}), GFM dual-loop droop (\ref{gfm_dual_loops}), GFM PI (\ref{gfmpi}), and SRF-PLL GFL (\ref{gfl}) controls.}
    \label{fig:local_dyn}
  \end{figure*}

\subsubsection{Probing signal} 
The voltage imposed by the ac source is given by $V_g = |V_g|\sin(\omega_gt)$, where $|V_g|$ and $\omega_g$ denote the voltage magnitude and frequency.  To excite the system, we apply a perturbation with frequency $\omega_p$
\begin{subequations}
  \begin{align}
    |V_g| &= V^\star + A_V \sin(\omega_p t)\\
    \omega_g &= \omega_0 + A_\omega \sin(\omega_p t).
  \end{align}
\end{subequations}
Here $V^\star \in \mathbb{R}_{>0}$ and $\omega_0$ denote the nominal voltage magnitude and frequency, $A_V$ and $A_\omega$ denote the amplitudes of the perturbations. To recover the transfer functions \eqref{eq:busdyn} at discrete frequencies $\omega_p$,  for each $\omega_p$ two experiments are conducted that separately perturb voltage magnitude (i.e., $A_V \in \mathbb{R}_{>0}$ and $A_\omega = 0$) and frequency (i.e., $A_V=0$ and $A_\omega \in \mathbb{R}_{>0}$).

\subsubsection{Measurement of output signals at the CIG bus}
At the CIG bus, we measure the three-phase voltage $v_{\text{abc}}$, $p$, and $q$ (see Fig.~\ref{fig:experiment}) in a synchronous reference frame rotating at the nominal frequency, i.e., $\theta_0(t)=\omega_0 t$. Next, the voltage magnitude $V$ and phase angle $\theta$ relative to $\theta_0(t)$ are computed. The amplitude and phase shift of the oscillations in the signals $V$, $\theta$, $p$, and $q$ are recovered by computing their Fourier Series coefficients corresponding to $\omega_p$. This allows representing the phase shift and magnitude of the perturbations as phasors $\Delta V \in \mathbb{C}$, $\Delta \theta  \in \mathbb{C}$, $\Delta p \in \mathbb{C}$, and $\Delta q  \in \mathbb{C}$. Finally, we can compute the frequency $\Delta \omega$ by applying the phase shift and gain of a differentiator at frequency $\omega_p$, i.e., $\Delta \omega = j \omega_p \Delta \theta$. A critical feature of this approach is that the data obtained for analysis is experimentally accessible without \emph{any knowledge} of the internal hardware or controls of the converter.  
\subsubsection{Recovering the frequency response model}
The phasors of the experiments at every frequency are collected in vectors 
\begin{align*}
  Y(j \omega_p) &=     \begin{bmatrix}
    \Delta\omega_{\omega}(j\omega_p) & \Delta\omega_{V}(j\omega_p) \\ \Delta V_{\omega}(j\omega_p) & \Delta V_{V}(j\omega_p)
  \end{bmatrix},\\ U(j \omega_p)&=     \begin{bmatrix}
    \Delta p_{\omega}(j\omega_p) & \Delta P_{V}(j\omega_p) \\ \Delta Q_{\omega}(j\omega_p) & \Delta q_V(j\omega_p)
  \end{bmatrix},
\end{align*}
where the columns correspond to perturbations of the ac source frequency and voltage, respectively. The transfer matrix \eqref{eq:busdyn} at the given frequency $\omega_p$ can then be immediately recovered by solving the equation 
  \begin{align}\label{eq:busdynnum}
    Y(j \omega_P) =
    \begin{bmatrix}
      g_{\omega,p}(j\omega_p) & g_{\omega,q}(j\omega_p) \\ g_{V,p}(j \omega_p) & g_{V,q}(j\omega_p) 
    \end{bmatrix}
    U(j \omega_p).
  \end{align}
%

\subsection{Bode plots of CIG bus dynamics}\label{sec:bodevalid}
Next, we apply our approach in an EMT simulation of the system shown in Fig.~\ref{fig:experiment} with detailed sampled-data control implementations with a controller sampling frequency of $10~\mathrm{kHz}$ and an averaged model of a two-level VSC\footnote{This study focuses on perturbation frequencies $\omega_p$ of up to $100$~Hz and therefore neglect converter switching that is typically in the range of $>2$~kHz.}. The resulting Bode magnitude plots are shown in Fig.~\ref{fig:local_dyn}. Notably, we implemented GFM droop control with cascaded inner controls (see, e.g., \cite[Fig.~4b]{acpower}), GFM PI control with inner loops, GFM droop control without inner control loops, and standard SRF-PLL GFL control with droop (see \cite[Fig.~4a]{acpower}).

The results grant insight into the limitations of the assumptions posed in the theoretical derivations and analytical models, especially for higher perturbation frequencies. In the ideal case (e.g., inductive coupling and infinite bandwidth of the inner control loops), it is expected that the diagonal entries of \eqref{eq:busdynnum} correspond to the $P-f$ and $Q-V$ droop characteristics of the outer controls \eqref{eqn:gfm_c}, \eqref{eq:gfmpi}, and \eqref{eqn:gfl_c}, while the off-diagonal entries of \eqref{eq:busdynnum} are zero.  For lower frequencies (i.e., up to $1~$Hz), this is reflected in the numerical results.  

Without the inner control loops, the filter impedance results in an additional voltage drop between the converter switch terminal and the terminal voltage used to compute \eqref{eq:busdynnum}. Therefore, GFM droop without inner loops does not reflect the reference $Q-V$ dynamics even at low frequencies. Furthermore, the off-diagonal entries of \eqref{eq:busdynnum} are only negligible for low perturbation frequencies.

The plots match the analytical predictions in the medium frequency range ($1$ to $60~$Hz) for GFM droop and GFM PI with inner loops, e.g., showing the high-frequency roll-off in $|g_{\omega,p}(j\omega_p)|$ expected from \eqref{eqn:gfm_c} and \eqref{eq:gfmpi}. In contrast, GFL displays an increase in $|g_{\omega,p}(j\omega_p)|$ since $g_{\text{gfl}}(s)$ is improper.

In the high-frequency range (after $60~$Hz), the numerical results no longer coincide with \eqref{eqn:gfm_c}, \eqref{eq:gfmpi}, and \eqref{eqn:gfl_c}. The inner loops are no longer able to accurately track the fast oscillations which compromises the CIG's ability to track the reference dynamics \eqref{eqn:gfm_c}, \eqref{eq:gfmpi}, and \eqref{eqn:gfl_c}. Moreover, for all controls, the off-diagonal entries of \eqref{eq:busdynnum} are no longer negligible highlighting that filter circuit dynamics and cross-coupling cannot be neglected at around line frequency. These results show that the performance specifications proposed in Sec.~\ref{sec:performance} can only be imposed up to the bandwidth limits of the controls. Investigating the impact of control and modulation bandwidth limits is seen as an interesting topic for future work. 
\subsection{External Measurement vs. Internal Reference}
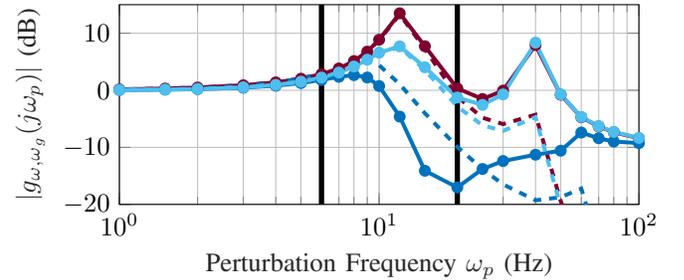
\begin{figure}[h!!!]
  \setlength\figH{0.3\linewidth}
  \setlength\figW{0.82\linewidth}
  \centering
%
%
\definecolor{mycolor1}{rgb}{0.3010 0.7450 0.9330}
\definecolor{mycolor2}{rgb}{0, 0.4470, 0.7410}
\definecolor{mycolor3}{rgb}{0.4980, 0, 0.1843}
\definecolor{mycolor4}{rgb}{0.0157, 0.7176, 0.5255}
\begin{tikzpicture}

\begin{axis}[%
width=0.951\figW,
height=\figH,
at={(0\figW,0\figH)},
scale only axis,
xmode=log,
xmin=1,
xmax=100,
xminorticks=true,
xlabel style={font=\color{white!15!black}},
xlabel={Perturbation Frequency $\omega_p$ (Hz)},
ymin=-20,
ymax=15,
ylabel style={font=\color{white!15!black}},
ylabel={$|g_{\omega, \omega_g}(j \omega_p)|$ (dB)},
axis background/.style={fill=white},
xmajorgrids,
xminorgrids,
ymajorgrids
]

\draw[draw=black, line width = 2pt] (6,-50) -- (6, 15);
\draw[draw=black, line width = 2pt] (20,-50) -- (20, 15);

\addplot [color=mycolor3, line width=1.5pt, mark size=1.5pt, mark=*, mark options={solid, mycolor3}, forget plot]
  table[row sep=crcr]{%
0.1	0.00169892546794263\\
0.15	0.00384003093071441\\
0.2	0.00685220109650512\\
0.3	0.0152745511105293\\
0.5	0.0414572602009762\\
0.8	0.100842779007698\\
1	0.151060142499521\\
1.5	0.301504314344755\\
2	0.475374756952225\\
3	0.880879493190598\\
4	1.38560399311325\\
5	2.02574954669574\\
6	2.82203187066025\\
7	3.82008786107261\\
8	5.11943341221182\\
9	6.76862108911184\\
10	8.86012578331237\\
12	13.4910362017069\\
15	7.71584900848596\\
20	0.325604940459418\\
25	-1.53098743386011\\
30	-0.0593674373886381\\
40	7.93656260548632\\
50	-0.779825488417273\\
60	-4.72426328617691\\
70	-6.33470321523978\\
80	-7.35222257261155\\
100	-8.41447363888644\\
};\label{gfmpi_ext}
\addplot [color=mycolor3, dashed, line width=1.5pt, forget plot]
  table[row sep=crcr]{%
0.1	0.00145566767003809\\
0.15	0.0038780498431311\\
0.2	0.00700185838254463\\
0.3	0.0154876890875411\\
0.5	0.0403768163415392\\
0.8	0.100105155982427\\
1	0.150899936786349\\
1.5	0.299309999714021\\
2	0.476132400835436\\
3	0.882524156160912\\
4	1.38724501979167\\
5	2.02614567375093\\
6	2.8182532266387\\
7	3.80836996329217\\
8	5.09416087311362\\
9	6.72270474038875\\
10	8.78341138143978\\
12	13.3150897759687\\
15	7.257092962315\\
20	-1.13188954394886\\
25	-4.81065355347499\\
30	-5.92621093485669\\
40	-4.28495760965269\\
50	-19.4055978368501\\
60	-29.1935609665873\\
70	-35.7766826412903\\
80	-41.238230556531\\
100	-49.763308517337\\
};
\addplot [color=mycolor2, line width=1.5pt, mark size=1.5pt, mark=*, mark options={solid, mycolor2}, forget plot]
  table[row sep=crcr]{%
0.1	0.000484126851641782\\
0.15	0.00108535132984009\\
0.2	0.00193146162334096\\
0.3	0.00434619804954489\\
0.5	0.0120767065717054\\
0.8	0.0309503463039328\\
1	0.0484083679080539\\
1.5	0.109295915041049\\
2	0.195217431075499\\
3	0.444715372661657\\
4	0.801919622685862\\
5	1.26655675909289\\
6	1.81593785109938\\
7	2.3604010678542\\
8	2.65419843245379\\
9	2.24170560264878\\
10	0.744952642632426\\
12	-4.61258244348782\\
15	-14.0994274817258\\
20	-17.0225625098663\\
25	-13.7918290276693\\
30	-12.4007491097759\\
40	-11.2925639692306\\
50	-10.5915234390348\\
60	-7.41425801496359\\
70	-8.40152482391431\\
80	-8.98864325619711\\
100	-9.25602518206457\\
};\label{gfm_noloops_ext}
\addplot [color=mycolor2, dashed, line width=1.5pt, forget plot]
  table[row sep=crcr]{%
0.1	0.000794495793320061\\
0.15	0.00178677107319089\\
0.2	0.00317815841331055\\
0.3	0.00715239352697446\\
0.5	0.0198736735057458\\
0.8	0.0509214924206247\\
1	0.079629275017458\\
1.5	0.179665914761042\\
2	0.320629098900648\\
3	0.7289079644858\\
4	1.31227318825671\\
5	2.07460030161151\\
6	2.99894220286897\\
7	4.00331102453887\\
8	4.85206119682563\\
9	5.10249989979324\\
10	4.39311938685771\\
12	1.07130492219093\\
15	-4.0221371909698\\
20	-9.92472556527946\\
25	-13.8085962840541\\
30	-16.4838686286561\\
40	-19.2858461559258\\
50	-18.7453442368399\\
60	-17.2218213032814\\
70	-27.2798866626384\\
80	-35.2541354044059\\
100	-45.9707992853579\\
};
\addplot [color=mycolor1, line width=1.5pt, mark size=1.5pt, mark=*, mark options={solid, mycolor1}, forget plot]
  table[row sep=crcr]{%
0.1	0.000549764444758345\\
0.15	0.00127305126770542\\
0.2	0.00228604503317912\\
0.3	0.00514178096349116\\
0.5	0.0142576533611256\\
0.8	0.0365898906419601\\
1	0.0570741915417762\\
1.5	0.129027740675908\\
2	0.230433532275809\\
3	0.523995218221644\\
4	0.946660864407117\\
5	1.50870221582305\\
6	2.22466010329276\\
7	3.10760778960493\\
8	4.16320055336353\\
9	5.36311330471428\\
10	6.58738295337429\\
12	7.70107240228098\\
15	4.03442864636669\\
20	-1.27840777640857\\
25	-2.55520654701318\\
30	-0.723998253944735\\
40	8.35318784768594\\
50	-0.7168551735095\\
60	-4.67449008124739\\
70	-6.2797323202561\\
80	-7.29341627319942\\
100	-8.35368167534407\\
};\label{gfm_loops_ext}
\addplot [color=mycolor1, dashed, line width=1.5pt, forget plot]
  table[row sep=crcr]{%
0.1	0.00163778533004493\\
0.15	0.00349710211393428\\
0.2	0.00507062160139396\\
0.3	0.00806766600996728\\
0.5	0.011847457214692\\
0.8	0.0349172258368748\\
1	0.0575087665441987\\
1.5	0.128869626484578\\
2	0.231887869627398\\
3	0.527760815842144\\
4	0.951796495360878\\
5	1.51471268464896\\
6	2.22838997428691\\
7	3.10559504175143\\
8	4.14895948515608\\
9	5.32896697403983\\
10	6.52235224852428\\
12	7.53207970738296\\
15	3.55854447382416\\
20	-2.85820288074993\\
25	-6.14583589664883\\
30	-7.11730456646244\\
40	-4.71452467905673\\
50	-20.351786849596\\
60	-30.2634667888607\\
70	-36.9243277149501\\
80	-42.4704375279541\\
100	-51.1870733055076\\
};
\end{axis}
\end{tikzpicture}%
  \caption{Bode magnitude plot of $g_{\omega, \omega_g}(j\omega_p)$ (solid) compared to that of $g_{\omega_{ref}, \omega_g}(j\omega_p)$ (dashed) for GFM single-loop droop (\ref{gfm_noloops_ext}), GFM dual-loop droop (\ref{gfm_dual_loops}), and GFM PI (\ref{gfmpi}).  The CIG bus frequencies diverge from the internal reference frequencies between $5~$Hz and $20~$Hz.\label{fig:ext_vs_int}}
\end{figure}

An advantage of the approach in Sec.~\ref{sec:sysid} over the approach in \cite{powertech} is that it captures the frequency dynamics at the CIG terminal while \cite{powertech} requires access to the internal frequency of GFM and GFL controls. To illustrate the differences between the frequency dynamics at the CIG terminal and reference frequency, we apply our approach from Sec.~\ref{sec:sysid} to both frequency signals. The results are compared in Fig.~\ref{fig:ext_vs_int}.  For GFM CIGs, the results from using the bus measurements show larger resonance peaks and less attenuation compared to those using the internal references.  This result shows that the internal reference frequency is not an adequate signal to assess the performance and interoperability of the response at the CIG terminal.

\subsection{Impact of coupling strength on GFL performance}\label{sec:SCRGFL}
\begin{figure}[h!]
  \setlength\figH{0.3\linewidth}
\setlength\figW{0.82\linewidth}
\centering
%
%
\definecolor{mycolor1}{rgb}{0.6196, 0.8706, 0.8000}
\definecolor{mycolor2}{rgb}{0.0157, 0.7176, 0.5255}%
\definecolor{mycolor3}{rgb}{0.00000, 0.4118, 0.2941}%
%
\begin{tikzpicture}

\begin{axis}[%
width=0.951\figW,
height=\figH,
at={(0\figW,0\figH)},
scale only axis,
xmode=log,
xmin=0.1,
xmax=100,
xminorticks=true,
xlabel style={font=\color{white!15!black}},
xlabel={Perturbation Frequency $\omega_p$ (Hz)},
ymin=-2.1,
ymax=0.2,
ylabel style={font=\color{white!15!black}},
ylabel={$|g_{\omega,\omega_g}(j \omega_p)|$ (dB)},
axis background/.style={fill=white},
xmajorgrids,
xminorgrids,
ymajorgrids
]
\addplot [color=mycolor1, line width=1.5pt, mark size=1.5pt, mark=*, mark options={solid, mycolor1}, forget plot]
  table[row sep=crcr]{%
0.1	-0.000897654354715126\\
0.15	-0.00201778338435678\\
0.2	-0.00358373240978838\\
0.3	-0.00803436753059465\\
0.5	-0.0220532045729383\\
0.8	-0.0548380926498507\\
1	-0.0834700843707815\\
1.5	-0.172387206874513\\
2	-0.275137353410504\\
3	-0.481808127310847\\
4	-0.660751781770431\\
5	-0.807844586702175\\
6	-0.928982353349889\\
7	-1.02924331380498\\
8	-1.11132662003804\\
9	-1.17729543724202\\
10	-1.22892488657693\\
12	-1.29666243646723\\
15	-1.33661594876457\\
20	-1.3248866833823\\
25	-1.28039580514969\\
30	-1.22787778907497\\
40	-1.11419365127241\\
50	-0.986947768634834\\
60	-0.877429011723243\\
70	-0.673876533092288\\
80	-0.482662375941504\\
100	-0.0202524679180632\\
  };\label{gfl_scr_2}
\addplot [color=mycolor2, line width=1.5pt, mark size=1.5pt, mark=*, mark options={solid, mycolor2}, forget plot]
  table[row sep=crcr]{%
0.1	-0.00103955335434805\\
0.15	-0.00233674709123739\\
0.2	-0.00414998159525831\\
0.3	-0.00930324912924317\\
0.5	-0.0255313281328264\\
0.8	-0.0634608390289533\\
1	-0.0965608125060533\\
1.5	-0.199209506500096\\
2	-0.317566082185061\\
3	-0.554839646518362\\
4	-0.759553191781558\\
5	-0.927603588134291\\
6	-1.06631768115291\\
7	-1.18191789976836\\
8	-1.27758284290996\\
9	-1.35561227428127\\
10	-1.41780033654366\\
12	-1.5020965958308\\
15	-1.55628962875953\\
20	-1.54920878114416\\
25	-1.4985790972501\\
30	-1.43710301757311\\
40	-1.30524785235447\\
50	-1.16081597281496\\
60	-1.08300463164289\\
70	-0.81063978209039\\
80	-0.597215848993918\\
100	-0.0770145729587372\\
};\label{gfl_scr_1}
\addplot [color=mycolor3, line width=1.5pt, mark size=1.5pt, mark=*, mark options={solid, mycolor3}, forget plot]
  table[row sep=crcr]{%
0.1	-0.0012917691122667\\
0.15	-0.00290363460206056\\
0.2	-0.00515628226571764\\
0.3	-0.0115576372369455\\
0.5	-0.0317058948472809\\
0.8	-0.0787399031605986\\
1	-0.119719536678222\\
1.5	-0.246425044512313\\
2	-0.391830363213231\\
3	-0.68122616718175\\
4	-0.928878086461654\\
5	-1.13124013770513\\
6	-1.29843376653018\\
7	-1.43891905106904\\
8	-1.55686851831853\\
9	-1.65518511004408\\
10	-1.73579242958991\\
12	-1.85116490427354\\
15	-1.93722322789434\\
20	-1.95130508225921\\
25	-1.89886923048418\\
30	-1.82688346696575\\
40	-1.66863326366439\\
50	-1.49812276342673\\
60	-1.4621193363516\\
70	-1.0925879010824\\
80	-0.846202186519883\\
100	-0.243151810390426\\
  };\label{gfl_scr_0.5}
\end{axis}
\end{tikzpicture}%
\caption{Bode magnitude plot of $g_{\omega, \omega_g}(j\omega_p)$ for SRF-PLL GFL control for a SCR of $2.5$ (\ref{gfl_scr_0.5}), $5$ (\ref{gfl_scr_1}), and $10$ (\ref{gfl_scr_2}).  \label{fig:gfl_scr}}
\end{figure}
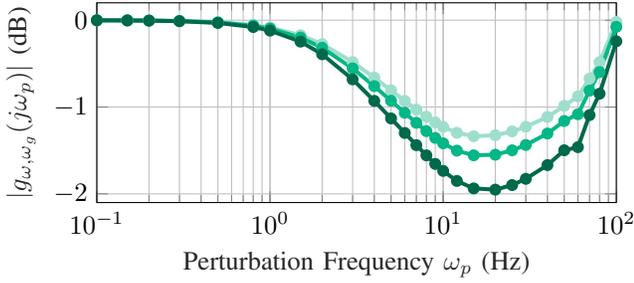

Numerical results for various grid coupling strength (i.e., SCRs) are shown in Fig.~\ref{fig:gfl_scr} and confirm the analytical results in Sec.~\ref{ex:gfl}, i.e., as the SCR increases $|g_{\omega,\omega_g}(j\omega_p)|$ tends to $0~$dB. This result indicates that under a stiffer network connection, the GFL CIG is better able to track the frequency of the infinite bus and keep synchronization for longer. In contrast, the minimum of $|g_{\omega,\omega_g}(j\omega_p)|$ lowers and the range of attenuation widens, as the SCR decreases.
%
%
\subsection{Impact of inner loop tuning on GFM droop control}
\begin{figure}[ht]
  \setlength\figH{0.3\linewidth}
  \setlength\figW{0.82\linewidth}
  \centering
%
%
\definecolor{mycolor1}{rgb}{0.3010 0.7450 0.9330}
\definecolor{mycolor2}{rgb}{0.6314, 0.9176, 1}
\definecolor{mycolor3}{rgb}{0, 0.4470, 0.7410}
\begin{tikzpicture}

\begin{axis}[%
width=0.951\figW,
height=\figH,
at={(0\figW,0\figH)},
scale only axis,
xmode=log,
xmin=1,
xmax=100,
xminorticks=true,
xlabel style={font=\color{white!15!black}},
xlabel={Perturbation Frequency $\omega_p$ (Hz)},
ymin=-20,
ymax=20,
ylabel style={font=\color{white!15!black}},
ylabel={$|g_{\omega, \omega_g}(j \omega_p)|$ (dB)},
axis background/.style={fill=white},
xmajorgrids,
xminorgrids,
ymajorgrids
]
\draw[draw=black, line width = 2pt] (60,-50) -- (60, 20);
\addplot [color=mycolor2, line width=1.5pt, mark size=1.5pt, mark=*, mark options={solid, mycolor2}, forget plot]
  table[row sep=crcr]{%
0.1	0.000491601823779234\\
0.15	0.000945385653263619\\
0.2	0.00167483381826884\\
0.3	0.00528932062134616\\
0.5	0.0144347817763205\\
0.8	0.0374260053693173\\
1	0.0571812331115876\\
1.5	0.131285435488987\\
2	0.235430379789403\\
3	0.532683599733522\\
4	0.965128432438766\\
5	1.53664553431309\\
6	2.26814117012786\\
7	3.16678560226721\\
8	4.25842732388763\\
9	5.52137239952401\\
10	6.93649134525461\\
12	9.61750166624599\\
15	10.5325727587709\\
20	13.4927268990025\\
25	16.2245123854085\\
30	5.83969445204259\\
40	-0.191673520690639\\
50	-2.25561055078788\\
60	-3.39323653264136\\
70	-3.77057934456687\\
80	-4.10630193319121\\
100	-4.48834997476145\\
};\label{gfm_low}
\addplot [color=mycolor1, line width=1.5pt, mark size=1.5pt, mark=*, mark options={solid, mycolor1}, forget plot]
  table[row sep=crcr]{%
0.1	0.000549764444758345\\
0.15	0.00127305126770542\\
0.2	0.00228604503317912\\
0.3	0.00514178096349116\\
0.5	0.0142576533611256\\
0.8	0.0365898906419601\\
1	0.0570741915417762\\
1.5	0.129027740675908\\
2	0.230433532275809\\
3	0.523995218221644\\
4	0.946660864407117\\
5	1.50870221582305\\
6	2.22466010329276\\
7	3.10760778960493\\
8	4.16320055336353\\
9	5.36311330471428\\
10	6.58738295337429\\
12	7.70107240228098\\
15	4.03442864636669\\
20	-1.27840777640857\\
25	-2.55520654701318\\
30	-0.723998253944735\\
40	8.35318784768594\\
50	-0.7168551735095\\
60	-4.67449008124739\\
70	-6.2797323202561\\
80	-7.29341627319942\\
100	-8.35368167534407\\
  };
\addplot [color=mycolor3, line width=1.5pt, mark size=1.5pt, mark=*, mark options={solid, mycolor3}, forget plot]
  table[row sep=crcr]{%
0.1	0.000484126851641782\\
0.15	0.00108535132984009\\
0.2	0.00193146162334096\\
0.3	0.00434619804954489\\
0.5	0.0120767065717054\\
0.8	0.0309503463039328\\
1	0.0484083679080539\\
1.5	0.109295915041049\\
2	0.195217431075499\\
3	0.444715372661657\\
4	0.801919622685862\\
5	1.26655675909289\\
6	1.81593785109938\\
7	2.3604010678542\\
8	2.65419843245379\\
9	2.24170560264878\\
10	0.744952642632426\\
12	-4.61258244348782\\
15	-14.0994274817258\\
20	-17.0225625098663\\
25	-13.7918290276693\\
30	-12.4007491097759\\
40	-11.2925639692306\\
50	-10.5915234390348\\
60	-7.41425801496359\\
70	-8.40152482391431\\
80	-8.98864325619711\\
100	-9.25602518206457\\
  };\label{leg:gfm_none}
\end{axis}
\end{tikzpicture}%
  \caption{Bode magnitude plot of $g_{\omega, \omega_g}(j\omega_p)$ for single-loop GFM control (\ref{leg:gfm_none}), low inner loop gains (\ref{gfm_low}), and high inner loop gains (\ref{gfm_dual_loops}).}
  \label{fig:gfm_tuning}
\end{figure}
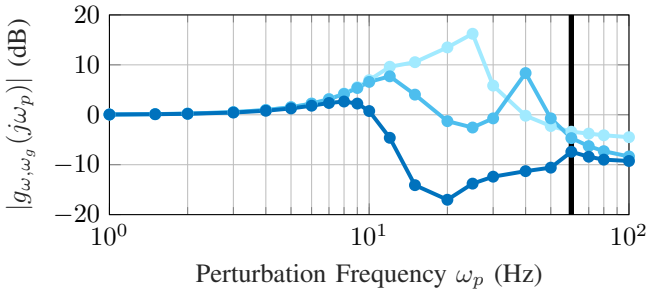

While the dual-loop GFM control structure is beneficial in tracking the GFM reference dynamics at the CIG terminal (see Fig.~\ref{fig:local_dyn}), it can also degrade the disturbance response if the inner loop gains are not tuned carefully. Figure.~\ref{fig:gfm_tuning} shows the transfer function $g_{\omega,\omega_g}(j\omega_p)$ for GFM droop control without inner loops, low inner loop gains, and high inner loop gains. The response without inner loops attenuates grid frequency disturbances in the range from $1$~Hz to $60$~Hz. Selecting inner loop gains on the stability boundary results in resonant peaks at $12$~Hz and $40$~Hz. In contrast, reducing the inner loop gains results in resonance at $25$~Hz. Thus, our method reveals that inner loop tuning results in a trade-off between tracking the reference dynamics and rejecting frequency disturbances. 

\subsection{Interoperability}\label{sec:numintop}
\begin{figure}[ht]
  \setlength\figH{0.4\linewidth}
  \setlength\figW{0.77\linewidth}
  \centering
  \input{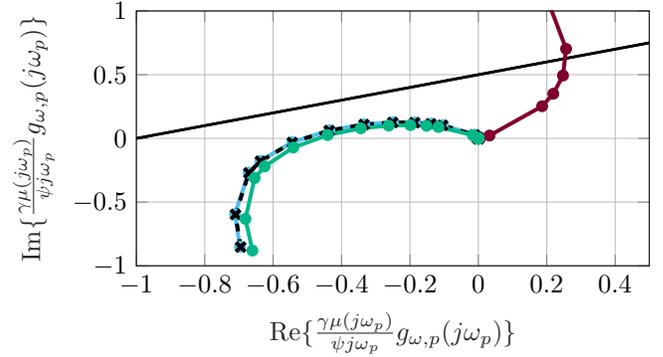}
  \caption{Nyquist diagram of $\tfrac{\gamma \mu(j\omega_p)}{\psi j\omega_p} g_{\omega,p}(j\omega_p)$ for GFM single-loop droop (\ref{dashed}), GFM dual-loop droop (\ref{gfm_dual_loops}), GFM PI (\ref{gfmpi}), and SRF-PLL GFL (\ref{gfl}) control considering  network circuit dynamics.}
  \label{fig:nyquist_num}
\end{figure}

Next, we illustrate that the interoperability conditions from Theorem~\ref{th.stability} can be verified using only input-output data. The Nyquist plot of $\tfrac{\gamma \mu(j\omega_p)}{\psi j\omega_p} g_{\omega,p}(j\omega_p)$ for various controls is shown in Fig.~\ref{fig:nyquist_num}. It can be seen that the results match the predictions based on analytical models in Sec.~\ref{ex:interoplinedyn}. Moreover, for the purpose of the stability conditions of Theorem~\ref{th.stability}, there is no significant difference between GFM droop control with and without inner loops. Finally, as discussed in Sec.~\ref{ex:interoplinedyn} the difference between GFM droop control and SRF-PLL GFL droop control is again small.

\subsection{Phase lag}
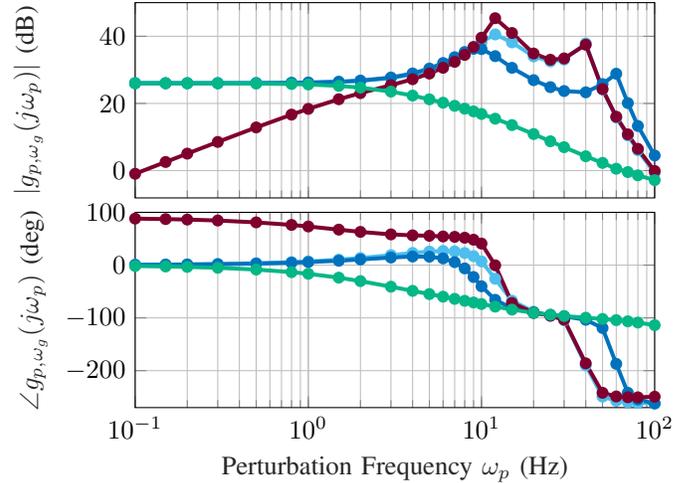
\begin{figure}[ht]
  \setlength\figH{0.7\linewidth}
  \setlength\figW{0.82\linewidth}
  \centering
%
%
\definecolor{mycolor1}{rgb}{0.3010 0.7450 0.9330}
\definecolor{mycolor2}{rgb}{0, 0.4470, 0.7410}
\definecolor{mycolor3}{rgb}{0.4980, 0, 0.1843}
\definecolor{mycolor4}{rgb}{0.0157, 0.7176, 0.5255}
\begin{tikzpicture}

\begin{axis}[%
width=0.951\figW,
height=0.419\figH,
at={(0\figW,0.45\figH)},
scale only axis,
xmode=log,
xmin=0.1,
xmax=100,
xminorticks=true,
xticklabels={,}
ymin=-5,
ymax=50,
ylabel style={font=\color{white!15!black},at={(-0.16,0.5)}},
ylabel={$|g_{p,\omega_g}(j \omega_p)|$ (dB)},
ylabel shift = -0.5 pt,
axis background/.style={fill=white},
xmajorgrids,
xminorgrids,
ymajorgrids
]
\addplot [color=mycolor1, line width=1.5pt, mark size=1.5pt, mark=*, mark options={solid, mycolor1}, forget plot]
  table[row sep=crcr]{%
0.1	26.0234651737219\\
0.15	26.026844671669\\
0.2	26.0304584163327\\
0.3	26.0395055443624\\
0.5	26.0625141419549\\
0.8	26.1321406867858\\
1	26.1968752668345\\
1.5	26.4128018607757\\
2	26.7102924017189\\
3	27.5172721827941\\
4	28.5695844312543\\
5	29.8254492367201\\
6	31.2594258837964\\
7	32.8572795861221\\
8	34.6066211818046\\
9	36.4683185109725\\
10	38.3153316187288\\
12	40.542261635088\\
15	38.1824521609698\\
20	33.9941798115191\\
25	32.5136428173771\\
30	33.0529091645069\\
40	37.8807829750906\\
50	24.1469285937397\\
60	15.799691256913\\
70	10.4662204259763\\
80	6.07240258064557\\
100	-0.715340738505709\\
};
\addplot [color=mycolor2, line width=1.5pt, mark size=1.5pt, mark=*, mark options={solid, mycolor2}, forget plot]
  table[row sep=crcr]{%
0.1	26.0215830366201\\
0.15	26.0240824724462\\
0.2	26.0275833795259\\
0.3	26.03757874698\\
0.5	26.0695114775574\\
0.8	26.1470344678146\\
1	26.2182054129879\\
1.5	26.4625411486748\\
2	26.7977924954229\\
3	27.717603531504\\
4	28.9288832872145\\
5	30.384482057995\\
6	32.0288601746193\\
7	33.7541094836649\\
8	35.3086436116702\\
9	36.2409582611766\\
10	36.1851157407776\\
12	34.080627581402\\
15	30.600871708095\\
20	26.9267666076833\\
25	24.8500290816499\\
30	23.6855520042766\\
40	23.3088082143408\\
50	25.7531408558283\\
60	28.8412533837027\\
70	20.110665130157\\
80	13.2894060521035\\
100	4.50393111023266\\
};
\addplot [color=mycolor3, line width=1.5pt, mark size=1.5pt, mark=*, mark options={solid, mycolor3}, forget plot]
  table[row sep=crcr]{%
0.1	-0.946868361173371\\
0.15	2.56527641430696\\
0.2	5.05063154195229\\
0.3	8.53427464570572\\
0.5	12.8529793636741\\
0.8	16.6679725031978\\
1	18.3815681774696\\
1.5	21.2504236731493\\
2	23.0663317027157\\
3	25.4301846494834\\
4	27.2109364150562\\
5	28.8896318764348\\
6	30.6089654654021\\
7	32.3792996656387\\
8	34.4445017524657\\
9	36.81215996733\\
10	39.5689871554637\\
12	45.3728066094326\\
15	40.9733808975978\\
20	34.8538834086157\\
25	32.999425148028\\
30	33.4025544301044\\
40	37.4772414188076\\
50	24.2642007137483\\
60	16.0430534172092\\
70	10.7890221261692\\
80	6.48106914806982\\
100	-0.114016200493332\\
};
\addplot [color=mycolor4, line width=1.5pt, mark size=1.5pt, mark=*, mark options={solid, mycolor4}, forget plot]
  table[row sep=crcr]{%
0.1	26.0157708342306\\
0.15	26.0109917248459\\
0.2	26.0042888897998\\
0.3	25.9851174579557\\
0.5	25.9240545895335\\
0.8	25.7785940242293\\
1	25.6487907083648\\
1.5	25.2285537175419\\
2	24.7075048023976\\
3	23.5241841150579\\
4	22.3301016768789\\
5	21.2202182508131\\
6	20.2102837012836\\
7	19.287004035281\\
8	18.4311916402693\\
9	17.6281996676377\\
10	16.8678285522709\\
12	15.4518336284325\\
15	13.5490356587708\\
20	10.8814101178144\\
25	8.72860175822926\\
30	6.97016341617589\\
40	4.27499843244865\\
50	2.29490815518741\\
60	0.556705269246585\\
70	-0.434401005407487\\
80	-1.40846328443169\\
100	-2.84693385786353\\
};
\end{axis}

\begin{axis}[%
width=0.951\figW,
height=0.419\figH,
at={(0\figW,0\figH)},
scale only axis,
xmode=log,
xmin=0.1,
xmax=100,
xminorticks=true,
xlabel style={font=\color{white!15!black}},
xlabel={Perturbation Frequency $\omega_p$ (Hz)},
ymin=-270,
ymax=100,
ylabel style={font=\color{white!15!black}},
ylabel={$\angle g_{p, \omega_g}(j \omega_p)$ (deg)},
axis background/.style={fill=white},
title style={font=\bfseries},
title={},
xmajorgrids,
xminorgrids,
ymajorgrids
]
\addplot [color=mycolor1, line width=1.5pt, mark size=1.5pt, mark=*, mark options={solid, mycolor1}, forget plot]
  table[row sep=crcr]{%
0.1	0.723421571511921\\
0.15	1.08090989901555\\
0.2	1.42108137757381\\
0.3	2.13813061865721\\
0.5	3.55480457351311\\
0.8	5.66031645575305\\
1	7.02512893696903\\
1.5	10.368607240972\\
2	13.4976190592994\\
3	18.9253091740928\\
4	23.0591372733161\\
5	25.7159566773361\\
6	26.767443381082\\
7	25.9846387029003\\
8	22.9630037873706\\
9	16.9885627122544\\
10	7.04336924899323\\
12	-25.8215763302366\\
15	-67.4386624479424\\
20	-88.1776659696845\\
25	-96.099765594135\\
30	-104.436337796564\\
40	-189.863413110668\\
50	-249.103129170956\\
60	-257.230844346773\\
70	-259.769554346013\\
80	-261.057025622974\\
100	-262.083238632662\\
};
\addplot [color=mycolor2, line width=1.5pt, mark size=1.5pt, mark=*, mark options={solid, mycolor2}, forget plot]
  table[row sep=crcr]{%
0.1	0.582534718794506\\
0.15	0.87356834362933\\
0.2	1.16434123980909\\
0.3	1.7446813162521\\
0.5	2.89807586025432\\
0.8	4.5991938126316\\
1	5.70577034686352\\
1.5	8.33681146914674\\
2	10.7120851334023\\
3	14.4201563334406\\
4	16.3389701274559\\
5	15.9854512637662\\
6	12.7243141836713\\
7	5.64593954620082\\
8	-6.2003198225708\\
9	-22.5088949007723\\
10	-40.0946864542322\\
12	-65.7135601773287\\
15	-81.9610668031172\\
20	-90.6917055079468\\
25	-94.3603590161932\\
30	-97.0628103888255\\
40	-103.553013083693\\
50	-119.193606166196\\
60	-187.111851877641\\
70	-241.864312628992\\
80	-254.414661168894\\
100	-262.209326293573\\
};
\addplot [color=mycolor3, line width=1.5pt, mark size=1.5pt, mark=*, mark options={solid, mycolor3}, forget plot]
  table[row sep=crcr]{%
0.1	88.1613312622961\\
0.15	87.2460895611731\\
0.2	86.3349290162627\\
0.3	84.5306665244303\\
0.5	81.0297173487502\\
0.8	76.1703906802554\\
1	73.2580167421978\\
1.5	67.2504748375556\\
2	62.9734420930988\\
3	58.245992892669\\
4	56.4271394155393\\
5	55.6138562792975\\
6	54.3096741515247\\
7	53.5800898851398\\
8	51.9149685742111\\
9	48.1369290824444\\
10	40.8608666580739\\
12	-0.347612955197775\\
15	-71.7502859999434\\
20	-90.0095476433528\\
25	-95.8301159435029\\
30	-103.299346567015\\
40	-186.156329738128\\
50	-241.923084454748\\
60	-248.891058923077\\
70	-250.401535402447\\
80	-250.535958002007\\
100	-249.340575571393\\
};
\addplot [color=mycolor4, line width=1.5pt, mark size=1.5pt, mark=*, mark options={solid, mycolor4}, forget plot]
  table[row sep=crcr]{%
0.1	-1.68971310369915\\
0.15	-2.53379908028674\\
0.2	-3.37686714241994\\
0.3	-5.0582865036954\\
0.5	-8.39045212112643\\
0.8	-13.2667243401746\\
1	-16.4063197243417\\
1.5	-23.7556285232106\\
2	-30.2873918210683\\
3	-40.8842855685379\\
4	-48.7844446891052\\
5	-54.8597468735128\\
6	-59.7787300588324\\
7	-63.9445056201634\\
8	-67.5886048056266\\
9	-70.8248401585606\\
10	-73.7191642864996\\
12	-78.6392520254672\\
15	-84.2594132500523\\
20	-90.3589039789682\\
25	-94.0522474590594\\
30	-96.4853005403742\\
40	-99.6879076855938\\
50	-102.086381651033\\
60	-104.237736715591\\
70	-106.50378678412\\
80	-108.814103351097\\
100	-113.764240734826\\
};
\end{axis}
\end{tikzpicture}%
  \caption{Bode plot of $g_{p,\omega_g}(j \omega_p)$ for GFM single-loop droop (\ref{gfm_noloops_ext}), GFM dual-loop droop (\ref{gfm_dual_loops}), GFM PI (\ref{gfmpi}), and GFL (\ref{gfl}) controls.  The phase plot shows a phase lag in the GFL response beginning at a much lower frequency than in the GFM responses.  \label{fig:grid_wp}}
\end{figure}

Finally, we note that different control implementations may be subject to various delays that manifest as phase shifts in the Bode plot and cannot be detected using Bode magnitude plots. Developing specifications for the Bode phase plots that rule out destabilizing phase lags is seen as an important direction for future work. Preliminary results for the phase shift of $g_{p,\omega_g}(j\omega_p)$ are shown in  Fig.~\ref{fig:grid_wp}. These results highlight that the SRF-PLL GFL frequency droop suffers from delays that cause phase lags in the Bode phase plot beginning at approximately $1$~Hz. In contrast, it can be seen that GFM control does not exhibit phase lag until around $6$~Hz.

\section{Conclusion}
In this work, we investigated interoperability and performance specifications for converter interfaced generation that can be verified with only input-output data. We first extended decentralized conditions for frequency stability to include network circuit dynamics. The conditions utilize a few key grid parameters and can be verified without exact knowledge of network parameters or topology. Moreover, we developed input-output performance specifications that characterize the disturbance response of CIG. Finally, we presented a data-driven validation method that allows for verification of the aforementioned interoperability and performance specifications using only input-output data. Moreover, the data-driven method was leveraged to illustrate the impact of key parameters such as inner control loop gains, network coupling strength, and controller bandwidth limitations. Extending the framework to performance specifications on phase plots, networks with heterogeneous R/X ratios, and GFM PI controls are seen as interesting areas for future work.

\bibliographystyle{IEEEtran}
\bibliography{IEEEabrv,bibliography}

\end{document}